\documentclass[aps,pre,twocolumn,showpacs,floatfix,preprintnumbers,superscriptaddress,amsmath,amssymb]{revtex4-1}
\usepackage{amsmath, amsthm, amsfonts, amssymb}
\usepackage{graphicx}
\usepackage{subfig}
\usepackage{natbib}
\usepackage{epstopdf}
\usepackage{caption}
\usepackage{footnote}
\usepackage{braket}
\input amssym.def
\usepackage{url}
\usepackage[colorlinks,bookmarks=false,citecolor=blue,linkcolor=blue,urlcolor=blue]{hyperref}
\begin{document}

\title{Atom interferometry using temporal Talbot effect on a Bose-Einstein condensate}
\author{Jay Mangaonkar}
\affiliation{Department of Physics, Indian Institute of Science Education and Research, Pune 411008, Maharashtra, India}
\author{Chetan Vishwakarma}
\affiliation{Department of Physics, Indian Institute of Science Education and Research, Pune 411008, Maharashtra, India}
\author{S. Sagar Maurya}
\affiliation{Department of Physics, Indian Institute of Science Education and Research, Pune 411008, Maharashtra, India}
\author{Sumit Sarkar}
\altaffiliation[Current address:] { SYRTE - Observatoire de Paris
61, avenue de l’Observatoire
75014 Paris - FRANCE}
\affiliation{Department of Physics, Indian Institute of Science Education and Research, Pune 411008, Maharashtra, India}
\author{Jamie L. MacLennan}
\affiliation{Department of Physics, University of Michigan, Ann Arbor, Michigan 48109, USA}
\author{Pranab Dutta}
\affiliation{Department of Physics, Indian Institute of Science Education and Research, Pune 411008, Maharashtra, India}
\author{Umakant D. Rapol}
\email {Electronic mail: umakant.rapol@iiserpune.ac.in}
\affiliation{Department of Physics, Indian Institute of Science Education and Research, Pune 411008, Maharashtra, India}

\begin{abstract}
We experimentally investigate a uniform pulse sequence in which atom interference is realized using the temporal matter-wave Talbot effect in an atom-optic kicked rotor system. Multi-path interference is obtained in a symmetric configuration with momentum differences up to $\pm$14 $\hbar k$, by virtue of Talbot resonance. We experimentally confirm the theoretical limit placed on the performance of this interferometer by the finite momentum distribution of the initial ensemble consisting of a Bose-Einstein condensate (BEC). This limitation on sensitivity, occurring due to the degradation of resonant dynamics is also important in the realization of a one dimensional continuous-time quantum walk in implementation of quantum search algorithms. 

\end{abstract}

\maketitle

\section{Introduction}
The optical Talbot effect is an interesting interference phenomenon occurring at sub-wavelength distances from diffracting elements. When a grating is illuminated by a plane wave of light, images having fractional grating periods are formed in the near field. The interference patterns revive at a characteristic length called the ``Talbot length" \cite{doi:10.1080/14786443608649032}. 
In the matter-wave analogue of this Talbot effect, similar revivals of the atomic wavefunction occur in the time domain for an Atom-Optic Kicked Rotor (AOKR) system \cite{PhysRevLett.83.5407}. In an AOKR system, an ensemble of cold atoms is subjected to a sequence of discrete pulses of a sinusoidal optical potential. This system has been shown to be a promising experimental test-bed for realizing quantum walks \cite{condmat5010004}, executing quantum search algorithms \cite{10.1088/1361-6455/ab63ad}, performing precision metrology \cite{PhysRevA.98.063614,PhysRevLett.89.224101,PhysRevLett.105.054103,PhysRevA.95.033601,PhysRevA.79.051402, PhysRevA.83.063613} and  probing fundamental physics \cite{PhysRevA.97.061601,doi:10.1002/andp.201600335,PhysRevLett.121.070402,White_2014,PhysRevE.83.046218}. 
The matter-wave Talbot effect that occurs due to quantum mechanical interference among the different momentum states can be used to realize a simultaneous multi-path atom interferometer \cite{McDowall_2009,PhysRevA.99.043617}. When the pulse period set to the Talbot time, Raman-Nath diffraction from the interaction with the optical grating leads to ballistic growth in the momentum imparted to the atomic ensemble \cite{PhysRevLett.96.160403}. These diffracted orders can then be made to interfere by tailoring an appropriate pulse sequence which leads to the revival of the initial state. The sensitivity of this revival depends on the number of participating orders in the interferometer. Ref. \cite{PhysRevLett.105.054103} is an example of such a scheme where a final phase-shifted high-order pulse was used to carry out this interference.

\begin{figure}[ht]
	\centering
	\includegraphics[width=0.7\linewidth]{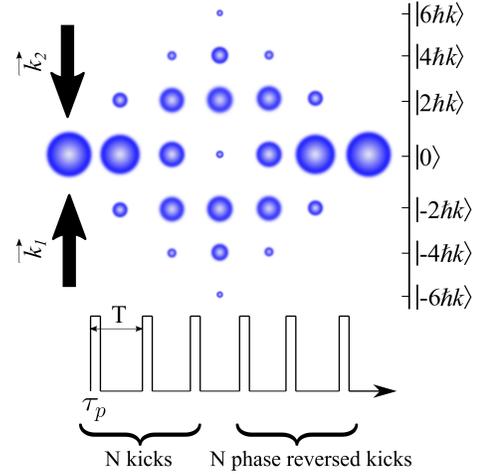}  
	\captionsetup{singlelinecheck=off,justification=raggedright}
	\caption{The evolution of the number of atoms in the BEC (number in $n^{th}$ order $\propto |\braket{2n \hbar k|\psi(p)}|^2$) is shown in momentum space at the resonance condition ($T=T_{T}$) for $\phi_{d}=0.7$. The spheres represent of the relative population present in different momentum states after each pulse in the sequence shown at the bottom. $\vec{k}_{1}$ and $\vec{k}_{2}$ denote the wave vectors of the two laser beams forming the standing wave that is pulsed according to the sequence shown. } \label{1}
\end{figure}

In this work, we experimentally realize a symmetric configuration for multi-path interference by applying a series of $2N$ kicks, where the second set of $N$ kicks are phase reversed for the retrieval of the initial wavefunction as proposed in Ref. \cite{PhysRevA.86.043604}.
The interfering atomic wavepackets coherently exchange momentum in integer multiples of 2$\hbar k$ ($k=2\pi$/$\lambda$) with the standing-wave optical field through a two-photon process. Thus, the basis set for this interaction is spanned by a one-dimensional grid of discrete momentum states separated by 2$\hbar k$. As shown in Fig. \ref{1}, when the pulse period is set to be equal to the Talbot time, the condensate wavefunction $\psi(p)$ after the first $N$ kicks is in a superposition of these basis states. The population in each $n^{th}$ state of the momentum basis set is given by $|\braket{2n \hbar k|\psi(p)}|^2$. As we show later, the free evolution phase term of the atomic wavefunction at the Talbot time is unity, and hence the only relevant phase is the one imparted by the standing wave pulses \cite{PhysRevLett.83.5407}. We change the sign of this phase midway through the sequence so that the phases imparted by the pulses mutually cancel at the end of the sequence. This results in retrieval of the initial momentum state and any extra phase gathered by the wavefunction will lead to an imperfect retrieval. The degree of retrieval of the initial state is captured by a quantity called fidelity. This quantity has been shown to be sensitive to the relative phases of the different interfering momentum wavepackets in Ref. \cite{McDowall_2009}. In this work, we measure the response of this fidelity to the deviation from the Talbot time. This measure helps us in understanding the factors affecting the sensitivity of such an interferometer, which plays an important role in its applicability towards metrology \cite{McDowall_2009,PhysRevA.83.063613,PhysRevE.87.020902} and quantum search algorithms \cite{10.1088/1361-6455/ab63ad}. 
We now discuss the interferometer in detail:  

\section{Theory}

 The dynamics of an atom with mass $m$ subjected to a periodically pulsed standing wave of far detuned laser light of wavelength $\lambda$ is governed by the following Hamiltonian of a standard kicked rotor (in scaled units):
 \begin{equation}
 \widehat{H}(t) = \hat{p}^{2} + \phi_{d} \cos(\hat{x}) \sum_{n=0}^{\infty} \delta(t-nl) 
 \end{equation}
 \begin{figure}[h]
	\centering
	\includegraphics[width=1\linewidth]{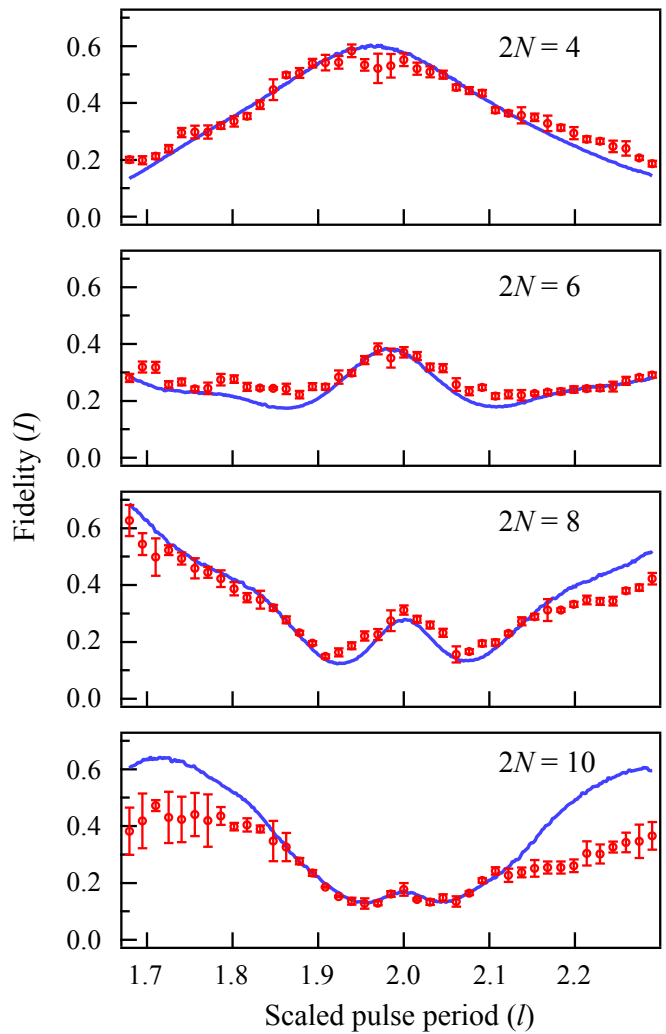}  
	\captionsetup{singlelinecheck=off,justification=raggedright}
	\caption{Fidelity ($I$) as a function of scaled pulse period ($l$) for different number of kicks $2N$ at a constant kick strength $\phi_{d}=0.8$. The red circles are the experimental data and the blue solid line is the numerical simulation. The error bars represent $\pm$ one standard deviation over 5 data points. $\Delta \beta$ is the only free parameter used to match the experimental values at different number of kicks. The values for $\Delta \beta$ used are 0.033, 0.027, 0.025 and 0.03 for $2N$=4, 6, 8 and 10 respectively. Data for $2N=2$ not shown here as fidelity doesn't undergo significant change at this scale.} \label{2}
\end{figure}
 
\noindent
This Hamiltonian has been extensively studied in the context of quantum chaos \cite{PhysRevLett.75.4598,PhysRevLett.118.174101,PhysRevE.100.060201}. The momentum $\hat{p}$ and position $\hat{x}$ are scaled in terms of the recoil momentum ($\hbar K$) and lattice spacing ($K^{-1}$) respectively, where $K=4\pi / \lambda$. The scaled period is $l=2T/ T_{T}$, where $T$ is the pulse period and $T_{T}= 4 \pi m / \hbar K^{2}$ is the Talbot time. The kick strength is given by $\phi_{d} = \Omega^{2} \Delta t / 8\delta$, where $\Delta t$ is the pulse duration and $\Omega$ is the effective Rabi frequency experienced by the atoms for a detuning of $\delta$ from the resonant transition. The momentum $\hat{p}$ can be expressed in terms of a discrete ($\hat{k}$) and a continuous component ($\beta$), such that $\hat{p}=\hat{k}+\hat{\beta}$, where $-1/2 < \beta \leq 1/2$. The interaction of the atoms with the periodically pulsed lattice changes their momentum state only in discrete units of $\hbar K$; $\beta$ for any initial momentum state is conserved. 

The time evolution of any initial state $\ket{k+\beta}$ can be obtained by applying a transformed Floquet operator \cite{PhysRevA.76.043415}. This operator $\widetilde{F}$ can be decomposed into two parts: the free evolution operator acting between two sequential kicks and the kick operator. Under the assumption that each kick can be approximated as a delta function ($T \gg \Delta t$), $\widetilde{F}$ can be written as- 

\begin{equation}
\widetilde{F}(\beta) = \left \{\exp \left[-i\pi l ({\hat{k} + \hat{\beta}})^2\right]\right \}\times \left \{\exp \left [i \phi_{d}\cos(\hat{x})\right ]\right \}  
\label{e2}
\end{equation}
The wavefunction after $N$ kicks is then determined by consecutive application of $\widetilde{F} (\beta)$.  For a kick sequence consisting of $2N$ kicks, the wavefunction of the final state can be written as :
\begin{equation}
\ket{\psi (t=2N)} = \left[\widetilde{F}_\pi (\beta)\right ]^N \left[\widetilde{F}_0 (\beta)\right ]^N \ket{k+\beta}    
\label{e3}
\end{equation}
where $\widetilde{F}_0 (\beta)$ is the Floquet operator for each of the first $N$ kicks and $\widetilde{F}_\pi (\beta)$ is that for the subsequent phase-shifted kicks. As mentioned previously, since we are interested in the retrieval of the initial wavefunction, we define  $I=  \mid$$ \bra{k+\beta}\widetilde{F}_\pi ^N (\beta) \widetilde{F}_0 ^N (\beta) \ket{k+\beta}$$\mid ^2 $, henceforth called as fidelity. 
For a plane wave initial state, $\ket{k+\beta = 0}$, at the resonance condition ($l=2$), the series of Floquet operators as given in Eq. $\ref{e3}$ collapse to unity, resulting in $I(l=2)=1$. As seen in Eq. $\ref{e2}$, for $l=2$, any deviation of the initial state by an amount $\beta$ results in a sub-optimal reversal as the free evolution phase can never become unity. Therefore, obtaining an optimal value of $I$ requires the initial distribution of atoms to have very narrow momentum distribution ($\Delta \beta$). Other major factors that affect the fidelity are noise in the phase and the amplitude of the standing wave lattice and in the timing noise of the kick sequence \cite{White_2014,PhysRevLett.118.174101}. 

An analytical expression for fidelity $I(l)$ has been derived in detail for an ideal plane-wave initial state in Ref. \cite{PhysRevA.86.043604}. We briefly present the results here. In the plane wave limit, i.e. ignoring quasi-momentum $\beta$, the fidelity becomes $I(l)=\mid$$ \bra{0}\widetilde{F}_\pi ^N \widetilde{F}_0 ^N \ket{0}$$\mid ^2$. As shown in Ref. \cite{PhysRevA.86.043604}, this leads to: 
\begin{equation}
I(l) = \left|  \sum_{n=-\infty}^{\infty} d_n ^* c_n \right| ^2   
\label{e4}
\end{equation}
where $c_n= \bra{n \hbar K} \widetilde{F}_0 ^N \ket{0}$ and $d_n ^* = \bra{n \hbar K} \widetilde{F}_\pi ^N \ket{0}^*$. Writing these complex coefficients in polar form and then using a first-order Taylor expansion in $l$, one arrives at their analytical forms:
\begin{equation}
c_n = J_n (N \phi_d) \exp [i \pi (A_+ (l-2) - n/2)] 
\label{e5}
\end{equation}
\begin{equation}
d_n ^* = J_n (N \phi_d) \exp [i \pi (A_- (l-2) + n/2)] 
\label{e6}
\end{equation}

\begin{equation}
	\begin{aligned}
		A_{\pm} & = \frac{1}{6} \left( N-\frac{1}{N} \right) n - \frac{1}{6} \phi_d (N^2 -1) \frac{J_{n-1}(N \phi_d)}{J_{n}(N \phi_d)} \\ 
		 & - \left( \frac{1}{3} N  \pm \frac{1}{2} + \frac{1}{6} \frac{1}{N} \right) n^2
	\end{aligned}
\label{e7}	
\end{equation}
where $J_n$ is an $n^{th}$ order Bessel function of the first kind.
The sum in Eq. $\ref{e4}$ can be  appropriately truncated to compute this value, as only finite orders are populated significantly during a pulse sequence. In the asymptotic limit of large $N$, a simple expression can be obtained by keeping only the dominant terms $\propto n^2 N$ in Eq. $\ref{e7}$. As shown in Ref. \cite{PhysRevA.86.043604}, Eq. $\ref{e4}$ can be reduced to:
\begin{equation}
I(l) \approx J_0^2 \left(\frac{\pi}{3} N^3 \phi_d ^2  (l-2) \right)   
\label{e8}
\end{equation}

The quantity of interest for the utility of this interferometry sequence is the sensitivity $S$ of the fidelity to deviations from resonance. It is defined as $S$ = ${\Delta l}/2$, where $\Delta l$ is the width of the fidelity peak $I(l)$ near the resonance condition $(l=2,T=T_{T})$. The pulse scheme we study here is attractive because of the rapid scaling it offers $S \propto N^{-3}$ with the number of kicks $N$, as evident in Eq. $\ref{e8}$. The scaling can be attributed to the significant relative phase change induced between the participating momentum orders at deviations of the pulse period from resonance \cite{McDowall_2009}.
In addition to the number of kicks, $S$ also scales as the inverse square of $\phi_{d}$ so that $S \sim 1/ (N^3 \phi_{d}^2)$. To improve the sensitivity, more diffraction orders should participate in the process, which requires either $\phi_{d}$ or $N$ to increase. Horne et. al. \cite{PhysRevA.83.063613} had investigated the performance limits of a class of similar kick sequences and had suggested parameter limits on $\phi_d$ and $N$, for which optimal sensitivity could be obtained. For all the sequences considered in that article, the finite momentum spread of the initial state was found to limit the maximum $\phi_{d}$ and $N$ that can be used without loss in sensitivity.

The deviation from the cubic scaling in sensitivity with $N$ can be understood by knowing the parameter range in which the assumptions Eq. $\ref{e8}$ makes fail to be valid. These assumptions are: 1. $N$ is such that the asymptotic approximation holds and 2. the initial state is a plane wave. For regimes where the first approximation is not valid, one can use Eq. $\ref{e4}$ to calculate $I(l)$. The validity of the second approximation of treating the BEC as an ideal plane wave can be checked as follows: In Ref. \cite{PhysRevA.86.043604}, the authors derive an expression for fidelity $F(l=2,\beta)$ for a plane wave initial state at the resonance condition, where $\beta$ is the deviation from the exact zero momentum eigenstate: 
\begin{equation}
F(l=2,\beta) \simeq J_{0}^{2} ( 4 \pi N^2 \phi_d   \beta  )
\label{e9}
\end{equation}
The width of $F(l=2,\beta)$ represents the range of values of $\beta$ which well-approximate the zero-momentum plane-wave state. As the number of pulses $N$ increases, the width of $F(l=2,\beta)$ is decreases inversely proportional to $N^2$. Deviations from numerical calculations of $I(l)$, that make the plane wave approximation are expected when the width of $F(l=2,\beta)$ is less than the momentum distribution of the initial state ($\Delta \beta$).

\section{Experiment}
We briefly present the experimental sequence here. A detailed description of the experimental set-up is provided in Ref. \cite{PhysRevLett.118.174101}. We obtain a  Bose-Einstein condensate (BEC) of \textsuperscript{87}Rb consisting of $\approx 3 \times 10^4 $ atoms after laser cooling in a standard Magneto-Optical Trap (MOT) and forced evaporative cooling in a hybrid 1064 nm crossed dipole trap. The trapping frequencies measured in the imaging plane are $2 \pi \times$(124$\pm6$) Hz and $2 \pi \times$(134$\pm7$) Hz. The temperature of the residual thermal component of the BEC is measured to be $79\pm13$ nK. The laser used for realizing the standing wave is locked to the $\ket{5 \ ^2 S_{1/2} , F=2 }$ $\rightarrow$ $\ket{5 \ ^2 P_{3/2} , F'=2 }$ D2 transition at 780 nm. Since the BEC is prepared in the $\ket{5 \ ^2 S_{1/2} , F = 1, m_{F} = -1}$ state, this laser is -6.8 GHz red-detuned from the atoms' accessible transition. A schematic of the experimental kick sequence is shown in Fig.\ \ref{1}. The kicking lattice whose phase can be arbitrarily controlled is arranged in a scheme similar to Ref. \cite{White_2014}. The pulse duration was set at $\approx$ 550 ns for this experiment and falls well within the criteria for the kicks to be in the Raman-Nath regime \cite{Gadway:09}. For these parameters, the probability of spontaneous emission per kick is $\sim$0.00142 which is low enough to be ignored. Similar to other BEC-based kicked-rotor experiments \cite{PhysRevE.83.046218}, the trapping fields are turned off about 100 $\mu s$ before the kicking pulses are delivered, to minimize the effect of mean-field interactions within the BEC. In order to spatially resolve the diffracted orders using time-of-flight detection, the atomic cloud is allowed to fall freely under gravity for 7 ms. 
From the absorption image, the fidelity is calculated as $I={f(0)}/{\sum f(n)}$, where, $f(n)$ denotes the number of atoms in the $n^{th}$ momentum state. 
Fig.\ \ref{2} shows the variation of fidelity $I$ vs. the kick period $T$, after application of the kick-sequence consisting of $2N$ kicks, for $N$=2$-$5. The kick strength for all sequences was $\approx$ 0.8. The fidelity plot has a maximum at the resonance condition of $l=2$ ($T=65.5 \ \mu s$) and the width of this resonance decreases with $N$.

\begin{figure}[ht]
	\centering
	\includegraphics[width=1\linewidth]{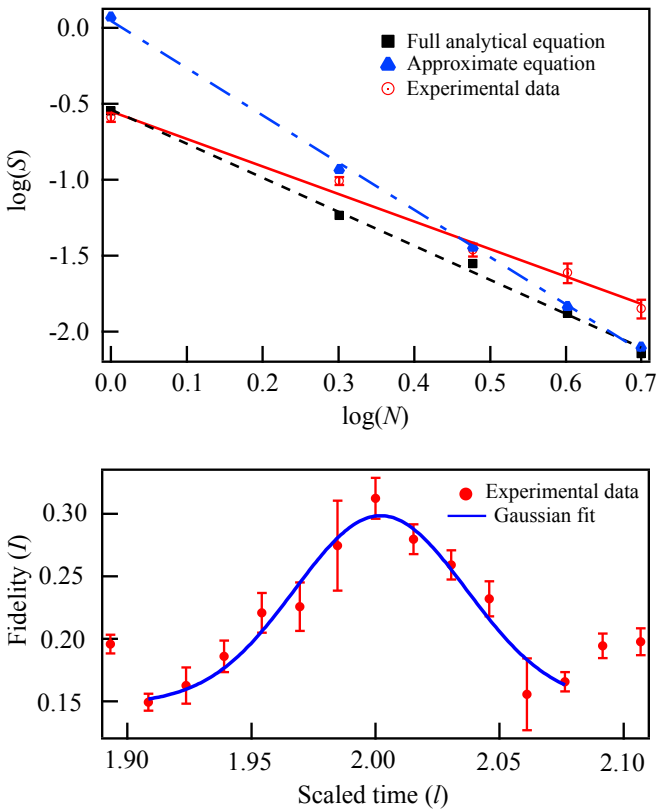}  
	\captionsetup{singlelinecheck=off,justification=raggedright}
	\caption{Top: Experimentally measured (red circles) and analytically calculated sensitivity $S$ as a function of number of pulses $N$, where $S=\Delta l / 2$ is the width of the fidelity $I(l)$ around resonance. The analytical equations used are Eq. $\ref{e4}$ (black squares) and Eq. $\ref{e8}$ (blue triangles). Solid line is a linear fit to experimental data. Dash and dash-dot lines are linear fits to $S$ determined using Eq. $\ref{e4}$ and Eq. $\ref{e8}$ respectively. Summation of Eq. $\ref{e4}$ was truncated to $n=\pm 20$ for calculation of $I(l)$. \\
	Bottom: Determination of $S$ from experimental data for $N=4$, by Gaussian fitting. The error bars are the standard deviation of 5 data points. Error bars in $S$ are $\pm$ standard deviation of the Gaussian width fit coefficient.} 
	\label{3}
\end{figure}
\section{Results}
To validate these experimental results, numerical simulations were performed in MATLAB with parameter values fitted to data but bounded within the estimated uncertainty of the experimental values. The numerical recipe was adapted from Ref. \cite{PhysRevA.76.043415}. In the simulation, we take the initial condition to be $1000$ non interacting plane waves in $k$-space which spans from $-20 \ \hbar K$ to $20 \ \hbar K$. Each $k$ bin is further divided into 101 parts to accommodate for the quasi-momentum $\beta$. The initial momentum state is constructed by drawing $p=k+\beta$ from a normal distribution $D(\beta)=\frac{1}{\Delta \beta \sqrt{2\pi}}\exp \frac{-(k+\beta)^2}{2 \Delta \beta^2}$, where $T_{\Delta \beta} =\frac{\Delta \beta^2 (\hbar K)^2}{m k_b} $ is the equivalent temperature of the initial state. The Floquet operators mentioned in Eq. \ref{e3} are then applied to evolve each state vector sample. Fidelity is then obtained by normalizing the number of states returning to the $k=0$ bin with the total number of states. This value is averaged over 10 simulation runs. While performing the experiment, the value of kick strength $\phi_d$ was fixed at $0.8 \pm 0.1$. Hence we keep $\phi_d$ = 0.8 for the simulations aswell. The experiment and simulation data are shown Fig. \ref{2}, where each experimental data point is an average of 5 runs and the error bars are $\pm$ one standard deviation in the measured fidelity. As one can see in Fig. \ref{2}, the best fit for experimental data was achieved for values of $\Delta \beta \sim 0.03$.  

To determine the experimental sensitivity $S$ of the pulse scheme, a Gaussian peak is fitted between the first minimas on either side of the central maxima of the experimental $I(l)$ curves (shown in Fig. $\ref{2}$) for each $N$ and the width ($\Delta l$) thus obtained is used to calculate $S = \Delta l / 2$. Fig. \ref{3} shows the variation of $S$ as a function of $N$. 
As shown in the figure we observe a scaling of $S$ $\propto$ $N^{-1.85 \pm 0.12}$ which is less than the expected scaling of $N^{-3}$ from Eq. $\ref{e4}$. Previous Talbot interferometry scheme with BEC had reported the exponent of this scaling to be $-2.73 \pm 0.19$ \cite{PhysRevLett.105.054103}. As explained in the theory section this discrepancy can happen due to two reasons: 1. asymptotic approximation is invalid and 2. the initial state as a plane wave is not a good approximation. As it can be seen in Fig. \ref{3}, for our experimental parameters i.e. for $\phi_d=0.8$ and $N=1-5$, Eq. $\ref{e8}$ deviates significantly from Eq. $\ref{e4}$ till $N=5$, after which the two converge. Thus the expected scaling exponent according to Eq. $\ref{e4}$, for $N<5$ is -2.23 which is still less than -1.85$\pm 0.12$ but larger than -3 as evident in Fig. $\ref{3}$. We now look at the approximation of the initial state as a plane wave. This approximation will hold in the regime where the width of initial wavepacket in momentum space is narrower than the width of $F(l=2,\beta)$ as calculated using Eq. $\ref{e9}$. For $\phi_d = 0.8$, the width of $F(l=2,\beta)$, at $N=2$ is 0.021. This is lower than the value of 0.03, which our simulations and experiment suggest, is the momentum width of our BEC. Thus we can expect the experimentally observed slowing down of the scaling of sensitivity with $N$ for $N \geq 2$.
For $N=1$ and $\phi_d = 0.8$, the width of $F(l=2,\beta)$ is $\sim$ 0.1. Since $\Delta \beta=0.03 < 0.1$, the plane wave approximation is valid. The value of $S$ calculated using Eq. $\ref{e4}$ for $N=1$ and $\phi_d=0.8$ is 0.285, which is close to the experimentally obtained value of $0.26 \pm 0.02$. These observations validate our model, where the finite momentum width of the initial state is responsible for the loss in the scaling of sensitivity $S$ with the number of pulses $N$.  

\begin{figure}[t]
	\centering
	\includegraphics[width=1\linewidth]{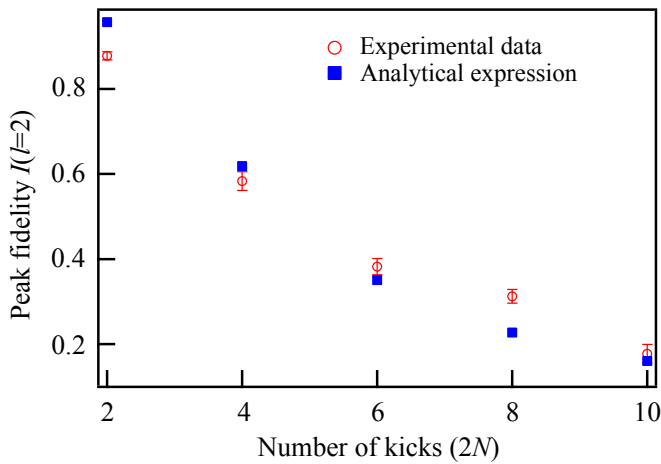}  
	\captionsetup{singlelinecheck=off,justification=raggedright}
	\caption{Peak fidelity at resonance $I(l=2)$ as a function of number of kicks $2N$. The red dots are experimental data of fidelity at resonance and the blue squares are theoretical values obtained from Eq. \ref{e9} for parameters $\phi_d = 0.8$ and $\Delta \beta=0.03$. Error bars on experimental values indicate $\pm$ one standard deviation over 5 data points. }\label{5}
\end{figure}

\begin{figure}[ht]
	\centering
	\includegraphics[width=1\linewidth]{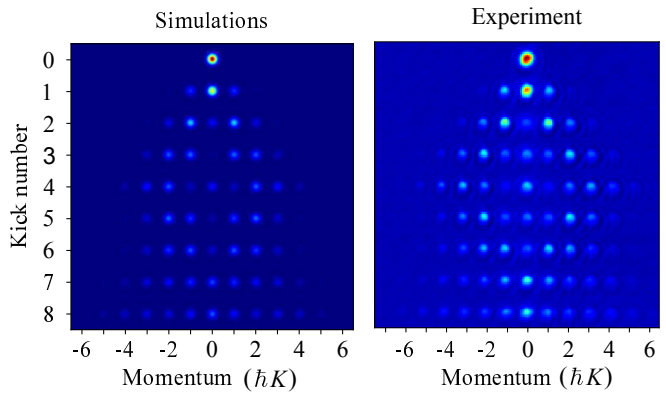}  
	\captionsetup{singlelinecheck=off,justification=raggedright}
	\caption{The evolution of population in different momentum states as a function of kick number at resonant pulse period $l=2$, where the phase of the lattice is inverted from kick 5 onward. The simulation data (left) depicts the amount of population in each momentum state for parameters $\phi_d = 0.8$, $2N=8$ and $\Delta \beta=0.03$. The experiment data (right) contains the absorption images taken for $\phi_d \sim 0.8$ after 7 ms time of flight.} \label{4}
\end{figure}

The suppression of fidelity at resonance ($l=2$) with $N$, can be calculated analytically using the distribution $D(\beta)$ and Eq. \ref{e9}. For a finite distribution initial state:
\begin{equation}
I(l=2) = \int_{-0.5}^{0.5} D(\beta) J_{0}^{2} ( 4 \pi N^2 \phi_d   \beta  ) d \beta
\label{e10}
\end{equation}
As seen in Fig. \ref{5}, the fidelity calculated using this expression agrees well with the experimentally observed value. 
The atoms which cause this suppression in fidelity because of finite $\beta$, leak to the other non zero momentum states $\ket{n \hbar K}$ as the phase shifted kick sequence is applied. 
This leaking of atoms to the non-zero momentum states, can be observed in the evolution of momentum distribution for a BEC as a function of kicks as shown in Fig.\ \ref{4}. In the left panel of the figure, each momentum order is depicted by a Gaussian distribution whose total area is proportional to the simulated population occupied by that order for $2N=8$, $\phi_{d}=0.8$ and $\Delta \beta = 0.03$. The right section shows the absorption images taken after each kick in the experiment for the same parameters. The population distribution observed experimentally resembles the distribution obtained from simulations. As suggested by the simulations, the performance of this scheme can be improved to some extent, by using a narrower momentum ensemble as the initial state e.g. ensembles having an order of magnitude lower $\Delta \beta$ than the one we utilize here have been reported ($\Delta \beta \leq 0.004$) \cite{PhysRevLett.96.160403,PhysRevLett.114.143004}. Thus at the current sensitivity level, which is majorly limited by temperature of the initial distribution ($\Delta \beta$), this multi-path interferometry scheme cannot compete with the state-of-the-art two-path interferometers as predicted in Ref. \cite{PhysRevA.83.063613}. 
As we discuss below, this limitation does not detract the utility of this demonstration in regards to the applications in quantum walker schemes and other BEC based AOKR experiments. 
In the context of the phase reversal sequence as a continuous-time quantum walk, the finite temperature effect needs to be considered while executing recently proposed schemes \cite{condmat4010010,10.1088/1361-6455/ab63ad,condmat5010004}. The quantum walk based search algorithm described in Ref. \cite{10.1088/1361-6455/ab63ad} relies on detecting atoms of a predefined tagged momentum state. However, such condition essentially demands the  reversal of wavepackets at non-tagged momentum states with high fidelity. As seen in Fig. \ref{4}, the large amount of population in the `off-target states'(non-zero momentum states) at the end of the sequence, reduces the amount of signal available for measurement of the `target state' (zero momentum state). For proposals of walks to determine topological phases \cite{PhysRevA.97.063603,condmat4010010}, the finite quasi-momentum acts like a phase noise and thereby compromising the exactness of the resonant condition on which the scheme relies on. This effect puts a limit on achieving the experimental  parameter regime where the desired signature can be observed. In Ref. \cite{condmat4010010} it was seen that the simulated signature of the topological phase becomes distorted for distributions $\Delta \beta \sim 0.03$. Thus the effect of finite momentum width on fidelity that we report, plays an important role in the above mentioned quantum walker dynamics.

\section{Conclusions} 
In conclusion, we have demonstrated an atom interferometer theoretically proposed in Ref. \cite{PhysRevA.86.043604}, using a $2N$ pulse sequence of a standing wave optical lattice. The optical lattice phase of the  first $N$ pulses and the remaining $N$ pulses differs by $\pi$ radians. We have probed the scaling of resonance width of fidelity with the number of pulses ($S \propto N^{a}$). The exponent of the observed scaling $a=-1.85\pm0.12$ differs from the analytically derived value of -2.23 due to the finite momentum distribution of the BEC. Simulations based on modeling the BEC as a distribution of finite momentum states are in good agreement with the observed data. These results show that in order to increase the performance of similar BEC based kicked rotor sequences, it is essential that the initial ensemble possess a very narrow momentum distribution. These observations of the effect of a finite momentum distribution are important for quantum simulations and quantum search algorithms based on the Talbot effect. 

\acknowledgments
The authors would like to thank the Council of Scientific and
Industrial Research (CSIR), Government of India, for funds through
Grant No. 03(1378)/16/EMR-II and the Indian Institute of Science Education and Research, Pune. CV would like to acknowledge CSIR for fellowship. JLM acknowledges support from NSF GROW (Grant No. DGE 1256260) and IUSSTF.

\end{document}